# Thermally irreversible sulfur chemisorption on silicate grains as a major sulfur reservoir in molecular clouds

Ni-En Sie[1], Masashi Tsuge[1,2], Yasuhiro Oba[1], Germán Molpeceres[3], Kenji Furuya[4], and Naoki Watanabe[1]

[1] Institute of Low Temperature Science, Hokkaido University, Sapporo, Hokkaido 060-0819, Japan, Corresponding author: niensie@lowtem.hokudai.ac.jp

[2] Department of Chemistry, Hiroshima University, Higashi-Hiroshima, Hiroshima 739-8526, Japan

[3] Departamento de Astrofísica Molecular, Instituto de Física Fundamental, CSIC, C/ Serrano 123, 113bis, 121, E-28006 Madrid, Spain

[4] RIKEN Pioneering Research Institute, 2-1 Hirosawa, Wako-shi, Saitama 351-0198, Japan

## Abstract

Sulfur (S) chemistry in the interstellar medium remains poorly constrained due to the long-standing "missing sulfur" problem, where a significant fraction of cosmic sulfur in molecular clouds remains unidentified. To gain a comprehensive understanding of S chemistry in the interstellar medium, its behavior on interstellar dust grains must be clarified. Here, we experimentally investigated the surface elementary processes of S atoms under conditions mimicking interstellar molecular clouds. Our in-situ laser-based detection method enabled the direct tracking of S atoms adsorbed on astrophysically relevant silicate substrates and amorphous solid water (ASW). We found that sulfur atoms strongly chemisorbs onto silicate surfaces and remains bound even after ASW sublimation, rendering it observationally invisible. An astrochemical model coupled with these results suggests that at least 10% of the total sulfur budget in molecular clouds must be locked onto silicate dust grains. Our results identify sulfur chemisorption on silicate as a key mechanism driving sulfur depletion and provide a physical explanation for the missing sulfur problem.

## 1. Introduction

Since the first detection of interstellar OCS and CS (Jefferts et al. 1971; Penzias et al. 1971), sulfur (S) chemistry has remained central to astrochemistry. Recent high-resolution observations with ALMA and JWST have revealed a rich S-bearing chemistry (Coulaud et al. 2025; Fuente et al. 2025; Taillard et al. 2025). However, while the cosmic abundance of S relative to H is $1.23 \times 10^{-5}$ (Snow & Witt 1996), detected S-bearing species in molecular clouds (MCs) account only for <5% of the total S budget (e.g., Tieftrunk et al. 1994; Jiménez-Escobar & Muñoz Caro 2011; Cazaux et al. 2022; Herath et al. 2025). This discrepancy, often referred to as the "missing sulfur" problem, suggests that sulfur is locked in solid phases on dust grains. A number of computational and experimental studies have been performed on this issue (e.g., Vidal et al. 2017; Laas & Caselli 2019; Yang et al. 2024; Slavicinska et al. 2025). The missing sulfur in MC may have been inherited and preserved in minerals like iron sulfides (FeS), magnesium sulfide (MgS) (Kama et al. 2019), identified in meteorites and in asteroid samples from Ryugu (Yokoyama et al. 2022) and Bennu (Lauretta et al. 2024).

Chemical models generally predict that sulfur initially accretes onto grain surfaces as atomic sulfur and then reacts on ice or dust in MCs (Garrod et al. 2007; Druard & Wakelam 2012; Vidal et al. 2017). In translucent clouds, $S^+$ rapidly accumulates onto dust grains (Ruffle et al. 1999; Vidal et al. 2017; Laas & Caselli 2019), where it is neutralized by electron transfer—a process feasible on silicates, polycyclic aromatic hydrocarbons (PAHs), and amorphous carbonaceous dusts. A widely discussed pathway is the sequential hydrogenation of atomic sulfur to $H_2S$ below 20 K (Garrod et al. 2007; Druard & Wakelam 2012; Vidal et al. 2017); however, solid $H_2S$ is constrained to <0.6% relative to $H_2O$ in MCs (McClure et al. 2023), possibly due in part to chemical desorption (Oba et al. 2018; Furuya et al. 2022) and low binding energy of $H_2S$ on $H_2O$ ice (Bariosco et al. 2024). Other proposed, yet to be confirmed, reservoirs include salts (e.g., $NH_4SH$) (Slavicinska et al. 2025), and sulfur allotropes (e.g., $S_8$), produced by irradiation-driven chemistry (e.g., Jiménez-Escobar & Muñoz Caro 2011; Cazaux et al. 2022; Carrascosa et al. 2024; Herath et al. 2025), while $S_8$ is the most stable form of sulfur allotropes and has been detected in meteorites and asteroid returned samples (e. g., Aponte et al. 2026). However, detecting such sulfur allotropes by astronomical observations remains challenging due to the lack of strong infrared features or permanent dipole moments (Taillard et al. 2025).

The behavior of S atoms on interstellar dust is governed by their binding energies (BEs). Theoretical models show a wide BE distribution on silicate and ice surfaces (Das et al. 2018; Perrero et al. 2022; Perrero et al. 2024a). On amorphous solid water (ASW), ground-state $S(^3P)$ is predicted to bind relatively weakly, with calculated BEs ranging from ~1500 K to 2800 K, whereas metastable $S(^1D)$ may chemisorb to form $S–OH_2$ or HOSH (Das et al. 2018; Perrero et al. 2022; Giustini et al. 2024; Perrero et al. 2024a; Di Genova et al. 2025). On silicates, calculations reported stronger adsorption, with BEs from 7,500 to 10,200 K on olivine nanoclusters through Mg sites and even higher values on amorphous silicate surfaces (Perrero et al. 2024b; Hansson et al. 2026). These calculated BEs lay near the typical chemisorption energy threshold of ~1 eV (~10,000 K) (Cazaux & Tielens 2004). Furthermore, the BEs of S allotropes ($S_n$, $n = 1–8$) on ASW show that $S_2$ generally exhibits a lower BE than atomic S, whereas larger allotropes ($S_n$, $n \geq 3$) bind more strongly as $n$ increases (Perrero et al. 2024a). These predictions make silicate grains plausible hidden sulfur reservoirs, but they require experimental validation at astrophysically realistic coverages.

Experimental constraints on S chemistry remain limited. Previous laboratory works have mainly examined photoprocessing of pure solid $H_2S$ and $H_2S$-containing ice mixtures using Fourier-transform infrared (FTIR) spectroscopy and temperature-programmed desorption (TPD) method, often with excessive sulfur concentrations (Jiménez-Escobar & Muñoz Caro 2011; Carrascosa et al. 2024). Although these experiments provided information for the possible production of S-bearing molecules, the missing sulfur problem has not yet been resolved. Moreover, the in-situ detection of atoms and radicals adsorbed on astronomical ice analogues is practically impossible by conventional experimental methods; hence, the behaviors of atoms/radicals on icy mantles or bare dust grain surfaces are not well understood (Tsuge & Watanabe 2023). To address this, we employed the PSD-REMPI method, previously used to study H, $H_2$, C, and OH on interstellar ice analogues (e.g., Watanabe & Tsuge 2020; Tsuge et al. 2023; Sie et al. 2024). We investigated the behavior of S atoms on amorphous silicate and ice surfaces over a range of temperatures to elucidate their fate in interstellar environments. Our measurements show that sulfur can remain strongly bound to silicate even after water-ice loss, supporting the idea that silicate grains constitute an observationally elusive reservoir of interstellar sulfur.

## 2. Experiments

### 2.1. Experimental setup and sample preparation

Experiments were conducted in an ultrahigh vacuum chamber (base pressure of $2 \times 10^{-7}$ Pa) equipped with a closed-cycle helium cryostat to cool an aluminum substrate to 10 K. To mimic interstellar grain surfaces, the substrate was coated with a 500 nm thick amorphous silicate ($Mg_2SiO_4$) film by radio-frequency sputtering (Kyodo International, Inc.). The morphological and compositional analyses of the silicate film produced in the same manner were performed in our previous work (Kouchi et al. 2021). For preparing the icy surface, a compact amorphous solid water (c-ASW) was formed on the silicate by background $H_2O$ vapor deposition at 100 K.

Since no dedicated S-atom source exists, $H_2S$ was used as the sulfur atoms precursor. Approximately 0.0025 monolayers (ML; 1 ML = $10^{15}$ molecules cm$^{-2}$) of $H_2S$ were deposited by background vapor deposition at 10 K on the icy/silicate surfaces. Because the target coverage is below the FTIR detection limit, the thickness was calibrated using deposition rates extrapolated from 0.1 ML $H_2S$. S atoms were produced *via* the photolysis of $H_2S$ at 30 K using an ArF excimer laser (193 nm, 10 min, flux ~$10^{15}$ cm$^{-2}$ s$^{-1}$ over a spot size of 2 cm$^2$, ExciStar XS, Coherent). This temperature ensured that H fragments thermally desorbed rapidly, preventing surface contamination. The decrease in the infrared (IR) absorption band of $H_2S$ (~2550 cm$^{-1}$) (Jiménez-Escobar & Muñoz Caro 2011) shown in Figure 1(a) indicated that most of the $H_2S$ was dissociated to form S atom. Based on the destruction rate, we confirmed that $H_2S$ on the surface would be dissociated in approximately 10 min.

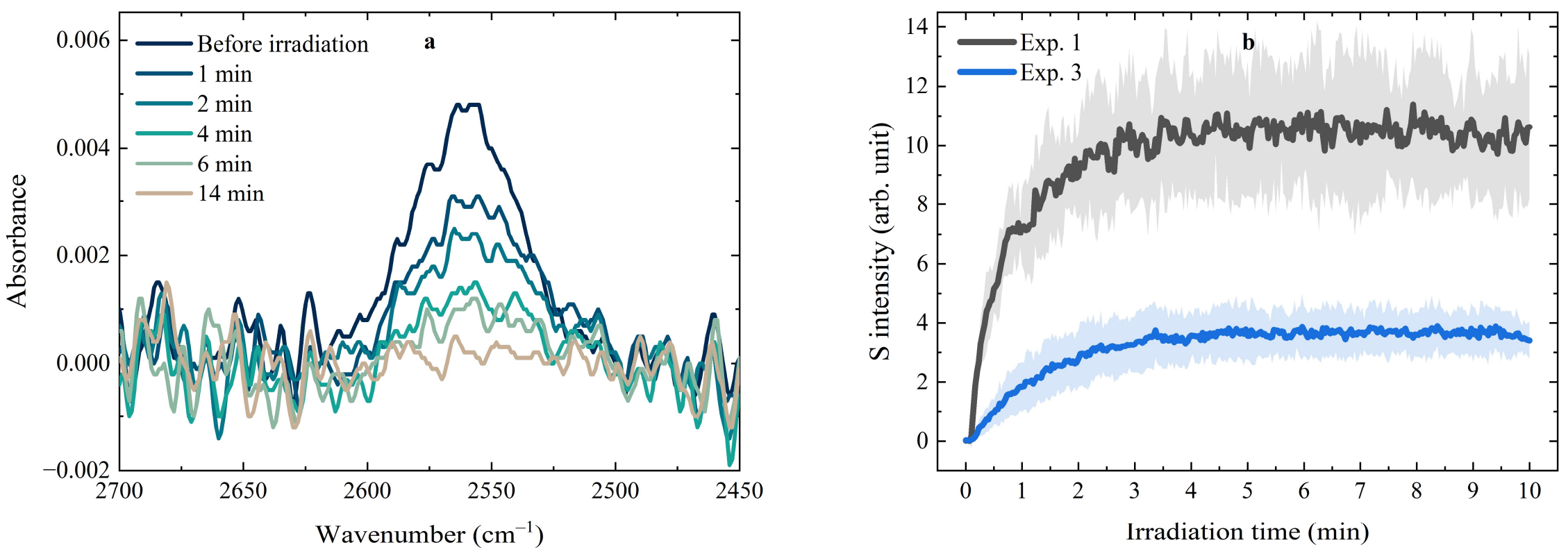

**Figure 1.** (a) The IR spectra of 0.1 ML $H_2S$ during 193 nm laser irradiation. (b) Substrate-dependent S-atom production curves. Increase of S-atom intensities during 193 nm photolysis on amorphous

silicate (Exp. 1, black) and c-ASW (Exp. 3, blue) at 30 K. Solid lines represent the average of several runs of experiments with shaded errors.

## 2.2. In-situ PSD-REMPI detection

S and $S_2$ were detected using a combination of photostimulated desorption and resonance-enhanced multiphoton ionization (the PSD-REMPI method), as schematically shown in Figure 2. The desorption was triggered by a pulsed nanosecond PSD laser (410 nm, 10 $\mu$J/pulse, 4 mm diameter) generated from an $Nd^{3+}$:YAG-pumped optical parametric oscillator (OPO). After a specific delay, the desorbed $S(^3P)$ atoms were selectively ionized above the surface at a typical distance of 2 mm *via* a (2+1) REMPI process (308–311 nm) provided by an $Nd^{3+}$:YAG-pumped dye laser (PrecisionScan, Sirah), targeting the $4^3P_J \leftarrow 3^3P_J$ transition (Brewer et al. 1982; Steadman & Baer 1988) and monitored at 308.115 nm (Figure 2). The $S_2$ products were detected at 316.163 nm *via* the $^3\Delta_g(3d\pi), v' = 0 \leftarrow X\,^3\Sigma_g^-, v'' = 0$ transition (313–317 nm) (Barnes et al. 1992). Delay time spectra were measured by changing the delay time between the PSD and REMPI laser pulses (delay time $= t_{REMPI} - t_{PSD}$). This profile corresponds to the translational energy distribution, and the integrated area of delay time spectra directly correlates with the surface number density of S atom (Sie et al. 2024). We confirmed the photodesorption is one-photon process by PSD laser power dependence experiments with a linear trend, excluding the desorption by phonon propagation (power-law dependence on the PSD power) or the thermal processes (exponential dependence on the PSD power). Due to the different detection efficiencies of the PSD-REMPI method, the number densities of S and $S_2$ cannot be directly compared.

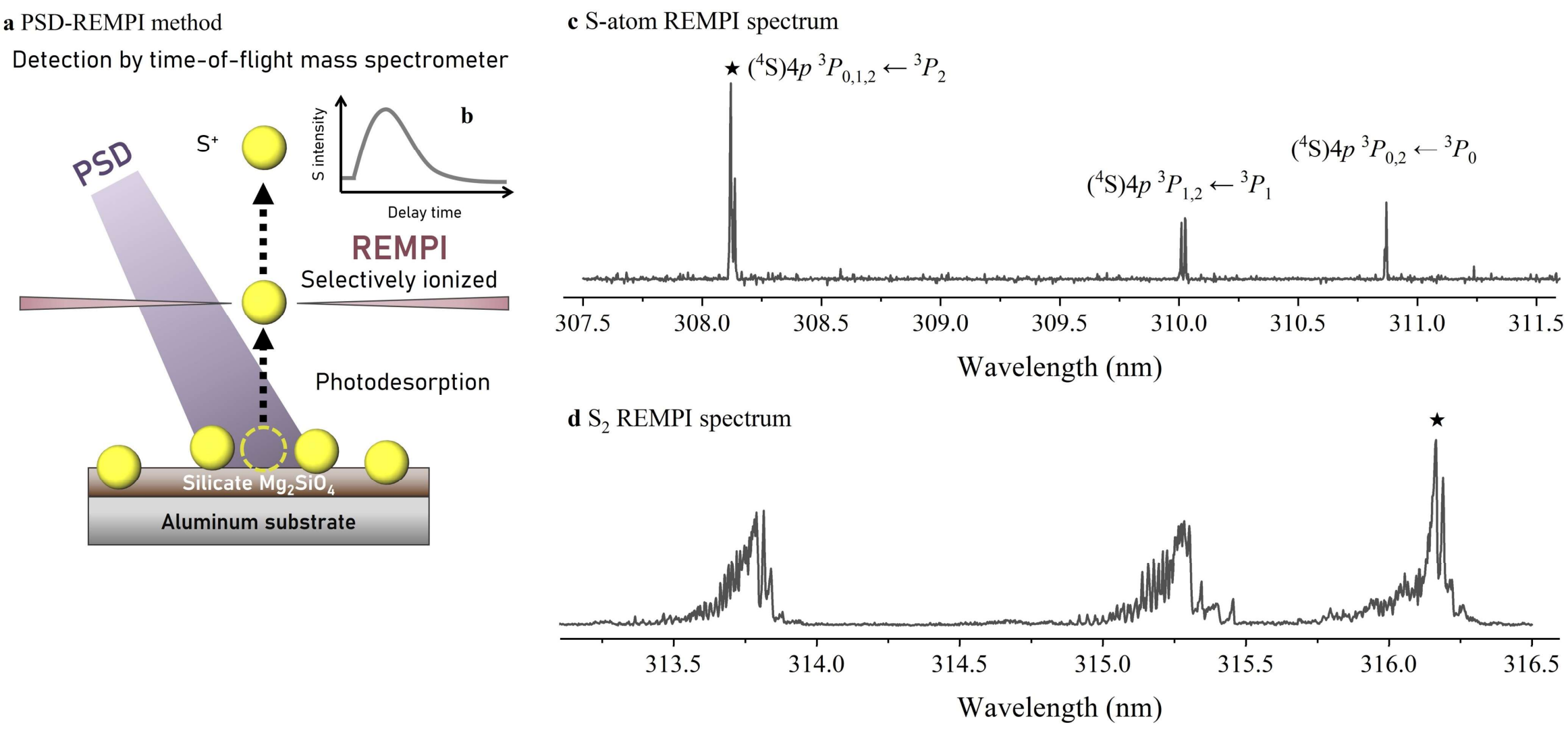

**Figure 2.** Left: (a) Principle of the in-situ PSD-REMPI method, showing the surface S atom is desorbed by a 410 nm pulsed laser and subsequently ionized by a REMPI laser. (b) Schematic of the delay time spectrum. Right: (2+1) REMPI spectra of (c) S atom and (d) $S_2$. Stars denote the peak wavelengths 308.115 nm and 316.163 nm used for tracing S atom and $S_2$, respectively.

## 2.3. Measurement protocol and sample configurations

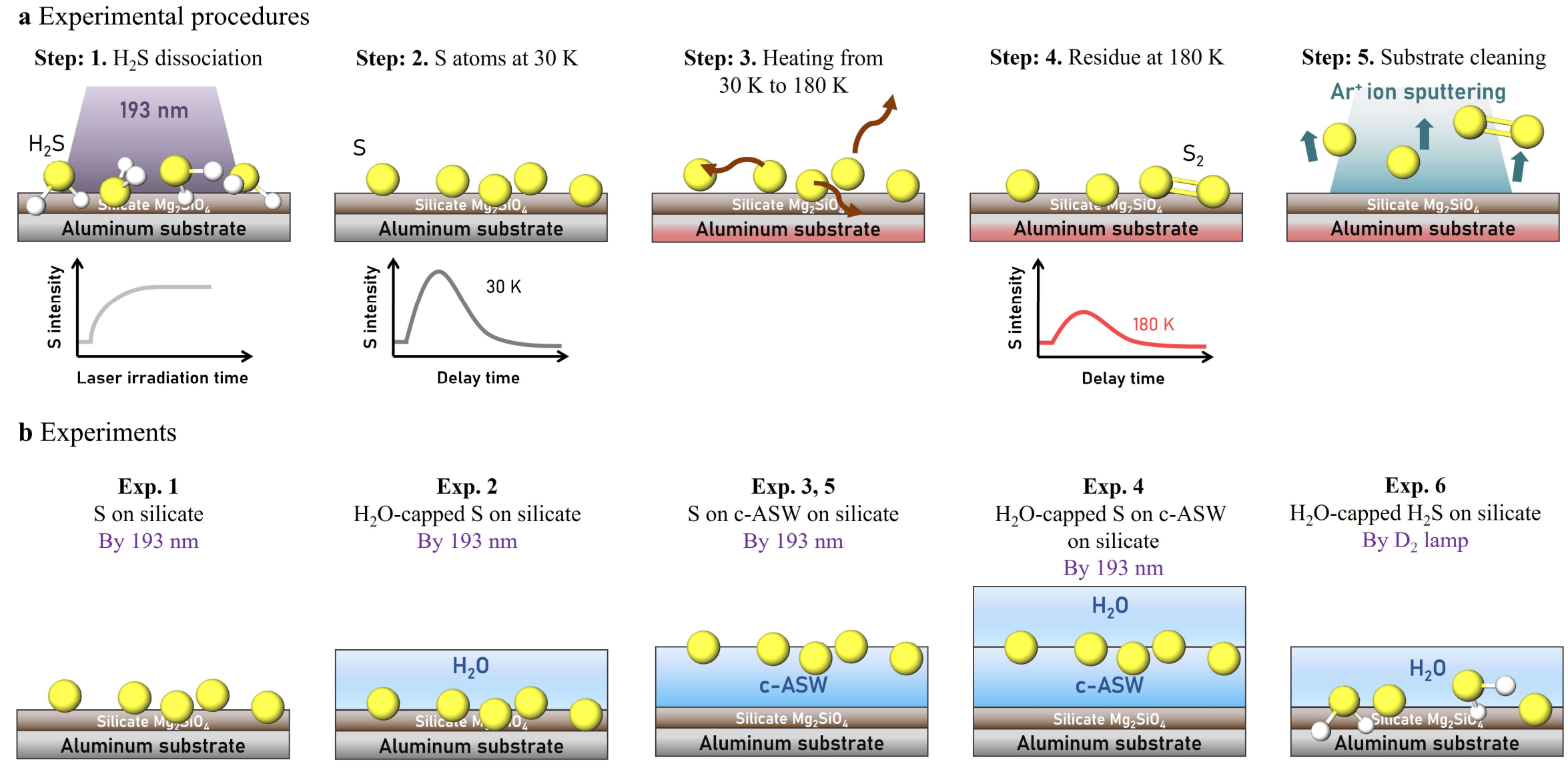

**Figure 3.** (a) Five-step experimental procedures sequence tracking sulfur production (30 K) and remnant (180 K) through thermal processing. (b) Structural schematics of six sample configurations.

Figure 3(a) illustrates the typical experimental five-step sequence: $H_2S$ photolysis at 193 nm to prepare S atoms on the surfaces (Step 1), measurement of the initial amount of S atoms at 30 K (Step 2), ramped heating at 10 K/min to 180 K (Step 3), the measurement of S atoms remaining at 180 K (Step 4), and substrate cleaning by $Ar^+$ ion sputtering (ion gun IQE 11/35, SPECS) to remove any residual $H_2S$ or photofragments (Step 5). Experiments with six different sample configurations were performed (Figure 3(b)): (1) S atoms adsorbed on bare silicate; (2) $H_2O$-capped S atoms on silicate; (3) S atoms on c-ASW-coated silicate; (4) $H_2O$-capped S atoms on c-ASW-coated silicate; (5) S atoms on c-ASW-coated silicate followed by stepwise heating; (6) $H_2O$-capped $H_2S$ on silicate irradiated by a $D_2$ lamp. For Exps. 2, 4, and 6, the water ice was additionally deposited on S atoms/$H_2S$ at 30 K. For Exp. 6, a $D_2$ lamp (115–400 nm, 30 min, flux ~$10^{13}$ photons $cm^{-2}$ $s^{-1}$, L11798, Hamamatsu Photonics K. K.) was utilized to simulate a cosmic-ray induced UV fluence (~ $2 \times 10^{16}$ photons $cm^{-2}$) equivalent to ~$10^5$ years of dense cloud exposure (Cecchi-Pestellini & Aiello 1992). The initial coverage of S atoms was identical across samples (1)-(5). In Exp. 5, the substrate was heated from 30 to 180 K in 10 K increments, with the intensities of surface S atoms recorded at each step. For each experiment, we defined the "S ratio" as the number density of S atoms remaining at 180 K relative to the initial number density at 30 K. These sample types and procedures are summarized in Table 1.

In Exps. 1 and 3, the S-atom signal was monitored in situ throughout the $H_2S$ photolysis (Figure 1(b)). Considering the photodissociation cross sections of both $H_2S$ and HS at 193 nm (~$5 \times 10^{-18}$ $cm^2$) (Heays et al. 2017), the saturated profiles indicate complete destruction of $H_2S$ into S atoms. The different intensities between Exps. 1 and 3 reflect that the PSD-REMPI detection efficiency is significantly higher on bare silicate than on c-ASW. The recombination between H and S atoms to form HS or $H_2S$ can be excluded at 30 K, as H atoms should thermally desorb immediately.

**Table 1. Summary of experimental parameters and sulfur retention ratios.**

| Exp. | Sample | Substrate | Irradiation source | S ratio[a] (180 K / 30 K) | $S_2$ ratio[a] (180 K / 30 K) |
|---|---|---|---|---|---|
| 1 | S | silicate | 193 nm laser | 0.50 ± 0.13 | 6.1 ± 3.4 |
| 2 | $H_2O$ capped S | silicate | 193 nm laser | 0.28 ± 0.16 | 8.3 ± 6.5 |
| 3 | S | c-ASW on silicate | 193 nm laser | 0.13 ± 0.01 | 8.7 ± 4.9[b] |
| 4 | $H_2O$ capped S | c-ASW on silicate | 193 nm laser | 0.14 ± 0.01 | 9.7 ± 2.0[b] |
| 5[c] | S | c-ASW on silicate | 193 nm laser | 0.11 ± 0.01 | N/A |
| 6 | $H_2O$ capped $H_2S$ | silicate | $D_2$ lamp | 0.10 ± 0.07 | 7.5 ± 3.0 |

**Notes.**

[a] S (or $S_2$) ratio represents the ratio of number density of S (or $S_2$) remained at 180 K relative to the initial number density at 30 K. The errors correspond to the standard deviation derived from multiple independent measurements.

[b] The PSD-REMPI detection efficiency for $S_2$ was assumed to be the same for silicate and c-ASW surfaces.

[c] Stepwise-heating experiment with 10 K increments.

# 3. Results

## 3.1. Chemisorption of S atoms on silicate

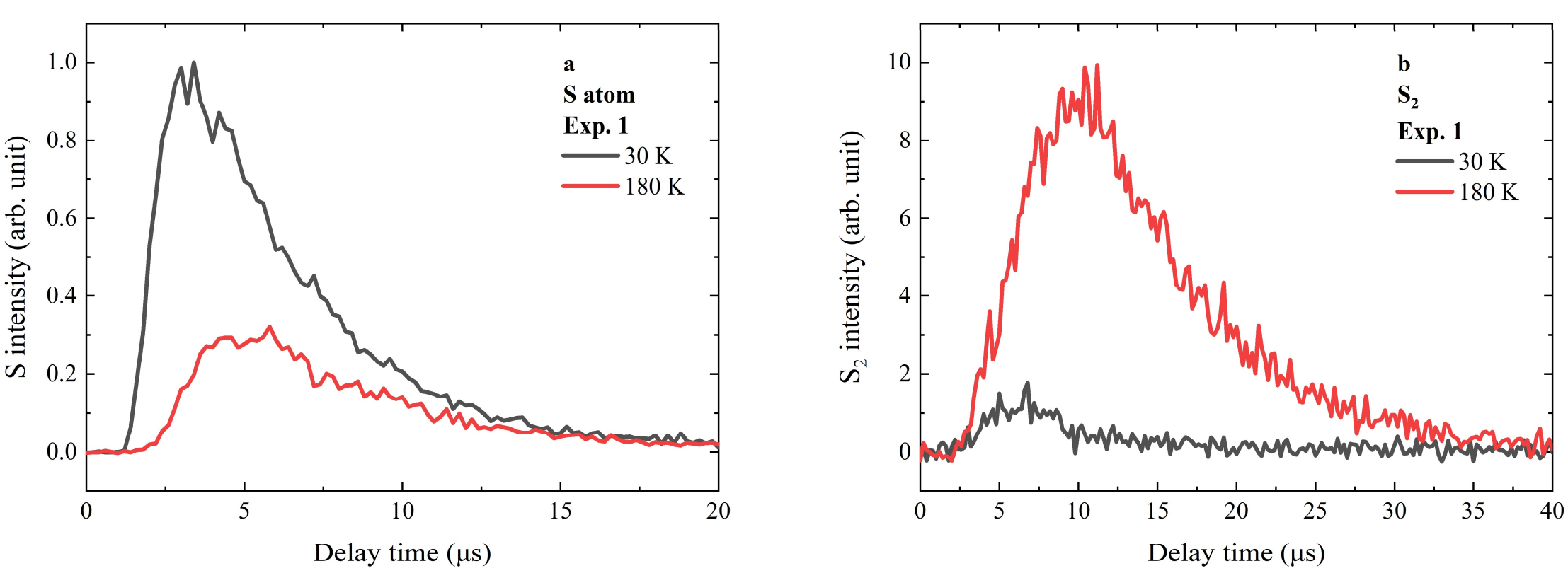

**Figure 4.** Delay time spectra of (a) S atom and (b) $S_2$ on amorphous silicate measured at 30 K (black) and 180 K (red) during Exp. 1.

We first investigated the behavior of S atoms on amorphous silicate surfaces (Exp. 1). Following 193 nm irradiation of $H_2S$ at 30 K (Figure 1(b)), the S-atom and $S_2$ intensities were measured as delay time spectra (Figure 4). The delay time ranges differ for S and $S_2$. The spectral profiles relate to desorption mechanisms, which are beyond the scope of this work. The delay time spectra of S and $S_2$ across all experiments show similar profiles, differing only in signal intensity. Therefore, presenting the data for Exp. 1 is sufficient to represent the overall trend.

Upon warming to 180 K, the S-atom intensity decreased significantly, while the $S_2$ intensity increased by a factor of 6 compared to an initial small amount produced at the preparation stage. This suggests that some fractions of S atoms were consumed in the diffusive formation of $S_2$ and probably larger allotropes. This is the first experimental evidence that sulfur cluster can form on silicate surface even when there is only a tiny amount of S atoms. Notably, PSD-REMPI intensities for both S and $S_2$ remained constant during subsequent annealing cycles between 180 K and 293 K. This refractory behavior indicates that S and $S_2$ are chemisorbed onto the $Mg_2SiO_4$ surface.

To ensure the origin of S and $S_2$ signals at 180 K are from chemisorbed themselves rather than the photodissociation of larger sulfur allotropes (e.g., $S_3$, $S_4$), several constraints were verified. Considering that the dissociation energy of $S_2$ is 4.37 eV (Okabe 1978), the 410 nm (3.02 eV) PSD photons are energetically insufficient to produce S atoms. While the dissociation of allotropes, such as $S_3$ (2.64 eV) (Francisco et al. 2005), may lead to form S atoms under PSD 410 nm. We repeated the PSD-REMPI measurements using a PSD wavelength of 600 nm, corresponding to a photon energy of 2.07 eV. This energy is below the reported gas-phase threshold of 2.64 eV for the dissociation of $S_3$ into S and $S_2$. The measurements using the 600 nm PSD laser showed behavior similar to that observed with 410 nm. The S signal exhibited a linear dependence on the 600 nm laser power, indicating that S was desorbed via a one-photon process. These observations support our assignment of the detected S signal to photodesorption of chemisorbed S rather than to photodissociation of larger S allotropes. Since the PSD-REMPI signals obtained with 600 nm PSD laser were much lower than that by 410 nm, in this work, we employed 410 nm for the PSD process.

Although there are no report on BE calculation for S on $Mg_2SiO_4$, the BE on Mg sites of $FeMgSiO_4$ is as large as ~7,500 to 10,200 K (Perrero et al. 2024b). Our quantum chemical calculations for a ground-state S-atom adsorption on forsterite nanocluster report a BE of approximately 9,700 K (~0.8 eV) when S atom is interacting with a single Mg site (Figure 5(a)), see Appendices A and B for more details), consistent with the observed refractory property. We propose that $S_2$ originates from the migration of physisorbed S atoms during the heating process before they transform into a chemisorbed state, a phenomenon previously reported for C atom on ASW (Tsuge et al. 2023). Once S atoms chemisorb on silicate, diffusive recombination to $S_2$ should be suppressed. The existence of refractory S-atom signals strongly indicates that the transformation from physisorbed- to chemisorbed-state happened during the heating process from 30 K to 180 K; however, we were unable to identify the exact temperature and the timescale of this transformation. This leads to difficulties in comparing the detection efficiencies for both physisorbed and chemisorbed S atoms in current experiments. The sublimation temperature of sulfur residue should be much higher than room temperature, which is unable to be examined under the current setup, and its determination is beyond the scope of this work.

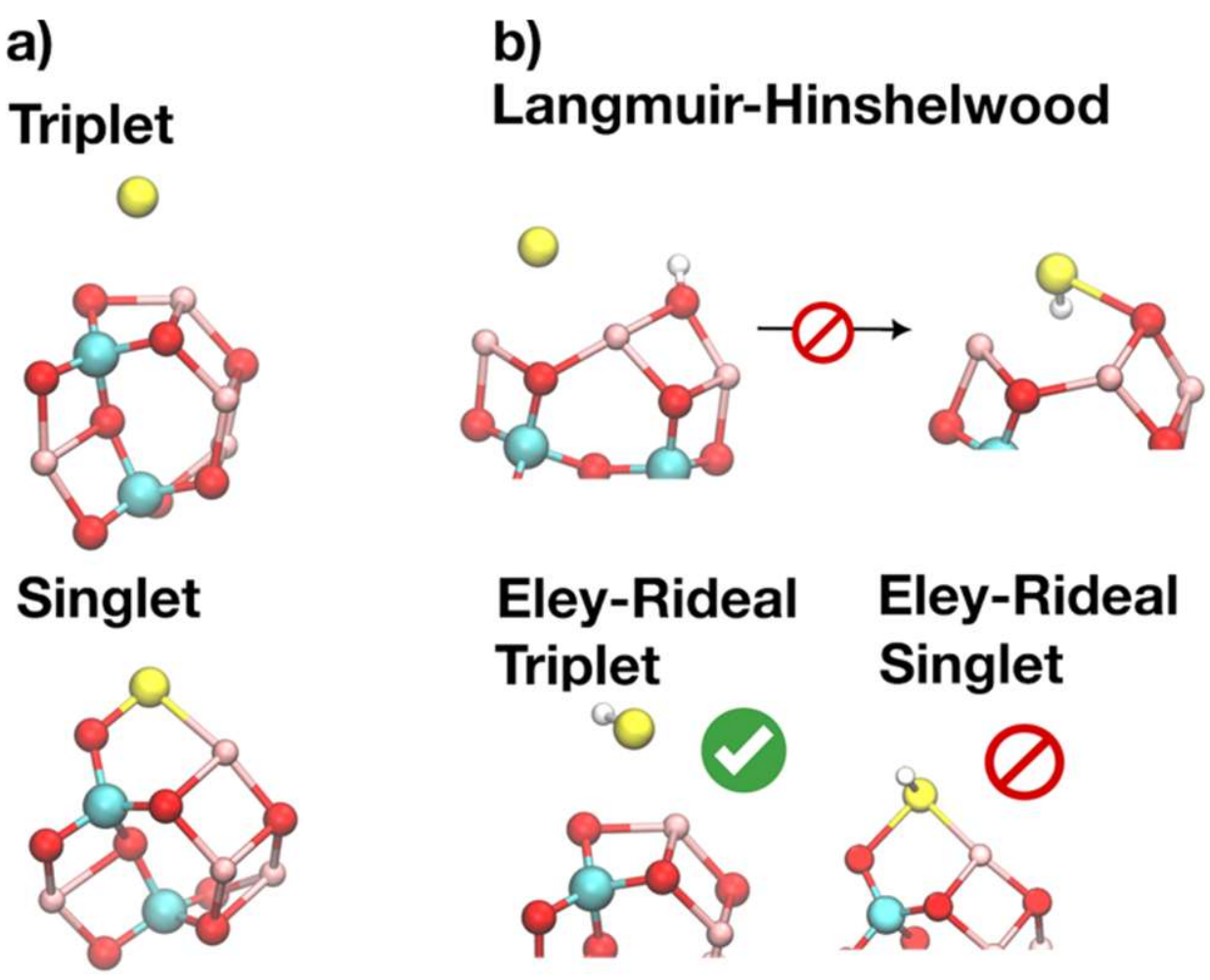

**Figure 5.** (a) Adsorption configurations of the triplet and singlet sulfur adsorption states on the forsterite nanocluster considered in both Perrero et al. (2024b) and the present work. (b) Schematic illustration of the hydrogenation pathways of atomic sulfur on silicates.

The S ratio was 0.50 ± 0.13 (Table 1). This reduction at 180 K is possibly attributed to: (1) the formation of $S_n$ allotropes, (2) a change in detection efficiency due to chemisorption, and (3) possible thermal desorption during warming up. The effect of (1) was confirmed by the detection of $S_2$. Although quantification is limited by the lack of relative detection efficiencies between physisorbed and chemisorbed sites in effect (2), the ratio can be compared with those determined in other experiments in the present study. The persistence of atomic S at room temperature confirms that silicate surfaces act as a stable sink for elemental sulfur. We will discuss effect (3) in the following sub-sections.

### 3.2. S atoms capped by a water ice layer

Since S atoms landing on silicate grain surfaces at the early stage in MCs would be capped by an amorphous $H_2O$ layer, we investigated the behavior of S atoms covered with an amorphous $H_2O$ layer (~40 ML; Exp. 2). This experiment would simulate the scenario suggested, for example, by Boogert et al. (2015). In MCs, $S^+$ is considered to land on a negatively charged grain surface to produce an S atom (Ruffle et al. 1999). In approximately $10^5$ years, the grain surface is expected to be covered with interstellar ices mainly composed of water. If we assume that the temperature is low enough, i.e., 10 K, during this period, S atoms are expected to stay at the interface between amorphous silicate and ice mantle. Finally, the sulfur-containing ice grains are thermally processed upon star formation. For this sample, we aimed to verify the fraction of S atoms that remains on grains after warm-up to 180 K, at which water ice has already been sublimated. Another practical purpose was that the thermal desorption of S atoms may be suppressed by the ice layer, which was considered as one of the reasons for S loss during warm-up processes (effect (3) mentioned above). In fact, no S-atom signals were detected from the ice surface during warming at temperatures below $H_2O$ sublimation. If thermal desorption of S atoms through ice layer could happen, the S atoms should appear on the top of the ice surface prior to the desorption event. After warming to 180 K, at which water ice has already been sublimated, the S ratio was found to be $0.28 \pm 0.16$, significantly lower than the fraction in the uncapped experiment (Exp. 1). However, the $S_2$ ratio was consistent within the analytical error between both experiments (Table 1). This indicates that the S-atom decline is due to neither mono-atomic thermal desorption nor an enhanced $S_2$ formation; instead, S atoms were likely co-desorbed with $H_2O$ during ice sublimation.

### 3.3. S atoms penetrating through ice mantle

Given that dust grain surfaces are generally covered with $H_2O$-dominated ices in MCs (Boogert et al. 2015), we examined S atoms deposited on a ~40 ML c-ASW formed on the silicate (Exp. 3). As seen in Figure 1(b), the detection efficiency of S atoms on silicate is 2.5 times larger than that on c-ASW. After correcting for this efficiency for initial S intensity on c-ASW at 30 K, the S ratio at 180 K was determined to be 0.13, substantially lower than on bare silicates. For $S_2$, the detected intensities were similar on silicate and c-ASW at both 30 and 180 K, suggesting that the PSD-REMPI detection efficiencies are comparable. Furthermore, the intensities for $S_2$ at 180 K were comparable to those in Exps. 1 and 2, indicating that $S_2$ formation was not suppressed by the underneath c-ASW (Table 1). The significant loss of S atoms at 180 K suggests that several competing processes govern the fate of S atoms on the c-ASW surface during warm-up: (1) thermal desorption from the c-ASW surface, (2) diffusive recombination into $S_2$, (3) co-desorption with $H_2O$, and (4) diffusion through the ice mantle to the silicate interface. To evaluate the pathway (1), the experiment with additional water ice on the top was performed (Exp. 4). The negligible impact of ice capping layers on S ratios (Exp. 4; Table 1) was found, indicating that thermal desorption from the ice surface is minimal. The observed $S_2$ supported the process (2), and low S ratios in Exps. 3 and 4 could be attributed to co-desorption with $H_2O$ as process (3).

To evaluate the pathway (4), we performed Exp. 5, where delay time spectra were measured with stepwise heating from 30 to 180 K in 10 K increments. Experiment 5 showed a striking trend: the S-atom signal on c-ASW decreased to nearly zero at 150 K but reappeared at 160 K, coinciding with $H_2O$ sublimation (Figure 6(a)). This "reappearance" indicates that S atoms penetrate the ice bulk during warming, eventually reaching the silicate-ice interface and chemisorbing onto the silicate surface. This is in line with the known thermal diffusion of physisorbed species through ASW (Watanabe & Kouchi 2008; Tsuge et al. 2020). In MCs, an ice mantle contains a variety of atoms and molecules; therefore, diffusion of S atoms into the ice may induce a rich **S** chemistry. Through these experiments, we suggest that, when S atoms are adsorbed atop an ice mantle, they can penetrate through it. Thus, the primary fate of S atoms on ice is either diffusive recombination into $S_n$ or migration to the silicate substrate. Moreover, if they reach the silicate surface, they will be chemisorbed and locked on the grain.

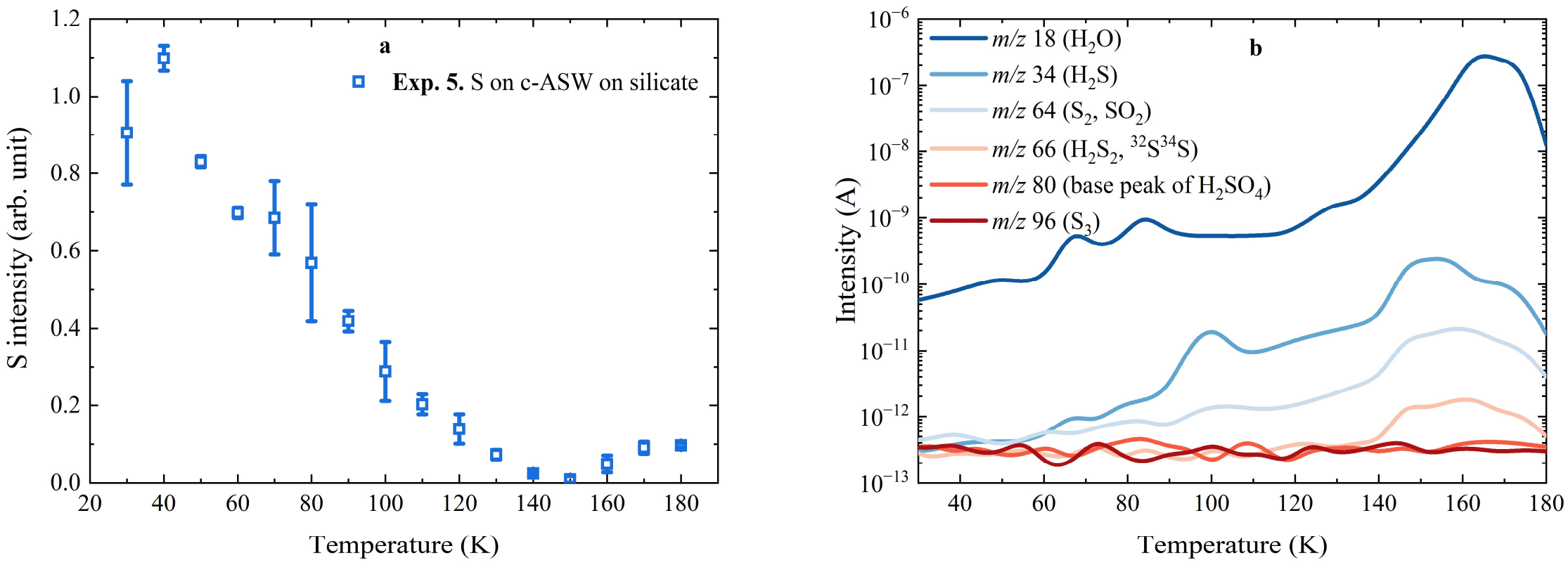

**Figure 6.** (a) S-atom intensity as a function of temperature on ice-covered silicate (Exp. 5). The intensity was measured at 10 K intervals, and error bars represent the deviation between two

independent measurements. (b) Temperature-programmed desorption of UV-irradiated $H_2O$-capped $H_2S$.

### 3.4. UV processing of $H_2O$-capped $H_2S$ on silicate

It has been generally assumed that S atoms on the silicate surface are first hydrogenated to yield $H_2S$ (Garrod et al. 2007; Druard & Wakelam 2012; Vidal et al. 2017), which would be further covered by $H_2O$-dominated ice mantles. In the inner part of dense clouds, the cosmic ray-induced secondary UV field (flux ≈ $10^4$ photons $cm^{-2}$ $s^{-1}$) (Cecchi-Pestellini & Aiello 1992) drives the chemistry of $H_2S$-containing ices, which was experimentally simulated previously (Jiménez-Escobar & Muñoz Caro 2011; Chen et al. 2014; Cazaux et al. 2022). However, the behavior of S atom and its allotropes remain to be clarified. Given that solid $H_2S$ has not been detected in the interstellar medium (ISM), while it is mainly formed on grains, it is likely that $H_2S$ is desorbed into the gas phase by chemical desorption (Oba et al. 2018) or converted to other S-bearing species, including "invisible" ones like S atoms on interstellar grains. Due to technical issues, the formation of such invisible S-bearing species has not been confirmed in situ at low temperatures (< 30 K) in former studies.

In Exp. 6, we photolyzed a small amount of $H_2S$ (~0.0025 ML, equivalent to Exps. 1 to 5) capped with 40 ML of water ice on silicate using a $D_2$ lamp to simulate the above conditions. Because the thick $H_2O$ overlayer prevents the direct detection of initial S-atom and $S_2$ intensity at 30 K, we performed a separate experiment with the absence of water layers to estimate the initial amount of S atom and $S_2$ by $D_2$ lamp photolysis of $H_2S$ on silicate at 30 K. By comparing the S-atom intensity measured at 180 K in Exp. 6 to this estimated initial value, we determined that approximately 10% of S atoms remained, and the $S_2$ yield at 180 K was obtained with an amount comparable to the other experiments (Table 1). Unlike irradiation at 193 nm, UV photons from the $D_2$ lamp can photolyze $H_2O$ as well as $H_2S$, so that the dissociated OH radicals and H atoms could have a chance to react with $H_2S$ and its photoproducts, HS and S atoms, leading to the formation of S-bearing molecules. The extremely low $H_2S$ coverage prevented IR detection. TPD measurements (Figure 6(b)) identified $S_2$ (*m/z* 64) and $H_2S_2$ (*m/z* 66) as photoproducts, while *m/z* 64 may contain contributions from $SO_2$, and *m/z* 66 also included signal from the $^{32}S^{34}S$ isotopomer. Other possible S-bearing products reported in previous studies, such as $S_3$ (*m/z* 96) and $H_2SO_4$ (base peak at *m/z* 80) (Jiménez-Escobar &

Muñoz Caro 2011; Carrascosa et al. 2024), were not detected partly because the number density is lower than the detection limit for the quadrupole mass spectrometer (M-201QA-TDM, Canon Anelva) in the present setup.

## 4. Astrophysical implications

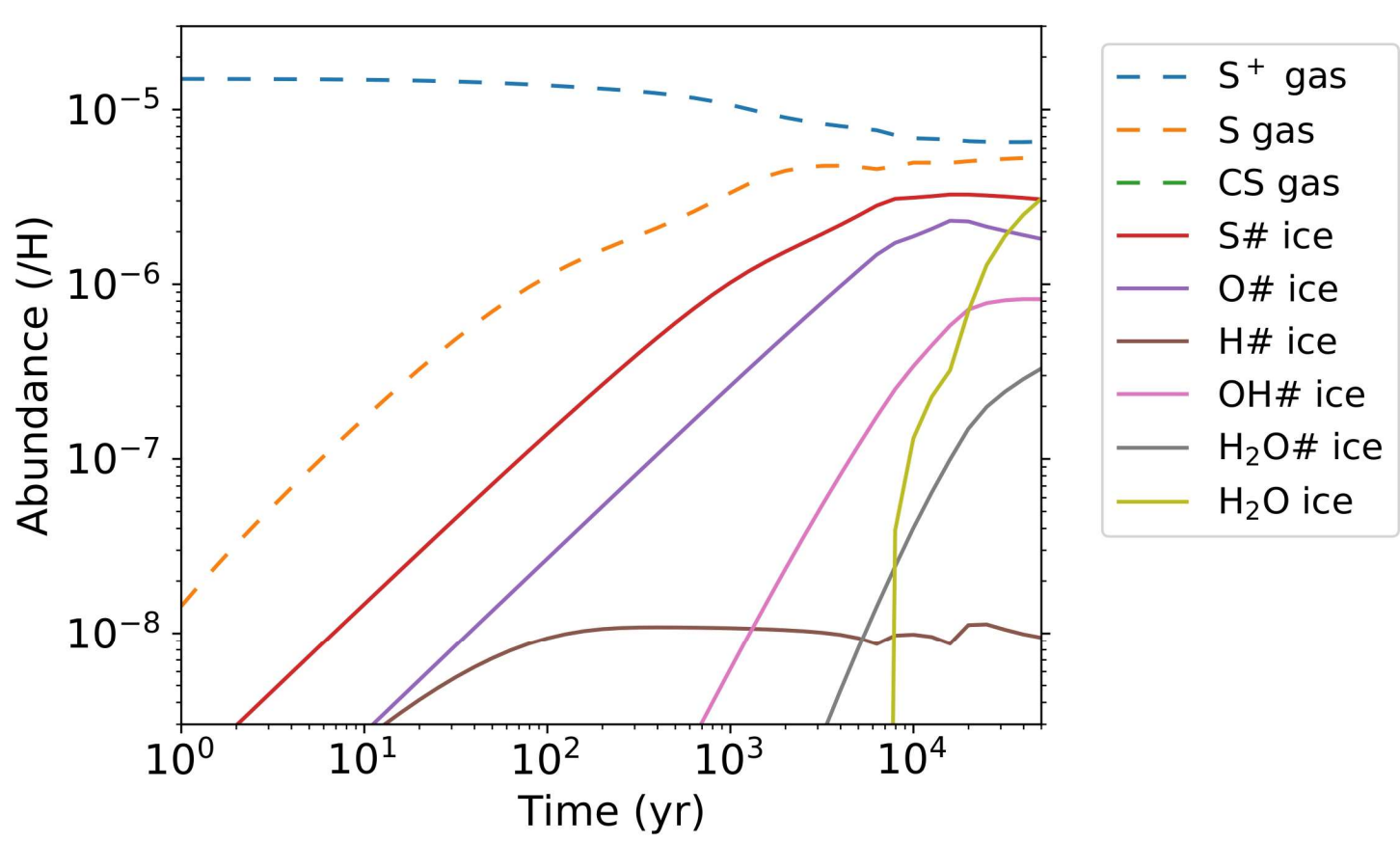

**Figure 7.** Astrochemical model under translucent cloud conditions (gas density of $10^3$ cm$^{-3}$ and Av = 1 mag) incorporating all our results. Time-dependent chemical abundances (relative to total H) of gas-phase and grain-surface species, where the '#' suffix denotes chemisorbed states.

We have simulated the evolution of a translucent cloud incorporating chemisorption on silicate grains and oxygen chemistry (Molpeceres et al. 2019). Details on the construction of the model can be found in Appendix C. Our model (Figure 7) shows a fast reduction of $S^+$ in timescales of a few $10^2$ years, simultaneously with a sharp rise in chemisorbed sulfur (S#). On the other hand, the abundance of chemisorbed O (O#) also increases but is lower than that of S# due to a higher ionization energy, making neutral oxygen the most abundant form of oxygen and reducing the adsorption rate on silicates. Most of $H_2O$ on the surface of bare grains ($H_2O$#) and chemisorbed OH (OH#) on silicate is formed through Eley–Rideal reactions. Once the silicate surface is mostly covered by S# and O#, the formation of $H_2O$ ice *via* Langmuir–Hinshelwood mechanism dominates, suppressing further sulfur entrapment. At this stage, chemisorbed sulfur species stay trapped below the proto-ice mantle and conventional sulfur chemistry, as investigated in other modelling works (e.g., Vidal et al. 2017; Laas & Caselli 2019).

Before this ice shielding effect occurs, our model estimates that 10–20% of the total cosmic sulfur abundance is stored on the grains as the chemisorbed sulfur species (that we include exclusively as S, based on our quantum chemical calculations shown in Appendices A and B), expanding the grain-locked sulfur reservoirs by a factor of 2–4 to what was previously known. These values, and the ultimate chemical form of sulfur remain subject to model uncertainties regarding surface binding site densities (Cuppen et al. 2017) and long-term radiolysis (Shingledecker et al. 2020).

These simulation results reframe sulfur evolution across the ISM, highlighting a critical evolutionary window during diffuse to dense cloud transition, as illustrated in Figure 8. The evidence for a potential elemental sulfur reservoir on bare exposed grains represents a step toward addressing the long-standing missing sulfur problem. In photodissociation regions (PDRs), sulfur is observable as $S^+$. During translucent clouds phase, neutralization occurs efficiently driven by Coulomb interactions with negatively charged grains and PAHs (van Dishoeck & Black 1989; Snow & McCall 2006). Once neutralized in the gas phase or on grains, atomic sulfur accretes irreversibly on bare grains with limited chemical conversion. At the same time, thick ice mantles will suppress the effect of sulfur chemisorption, at least at low temperatures, when S atoms cannot diffuse.

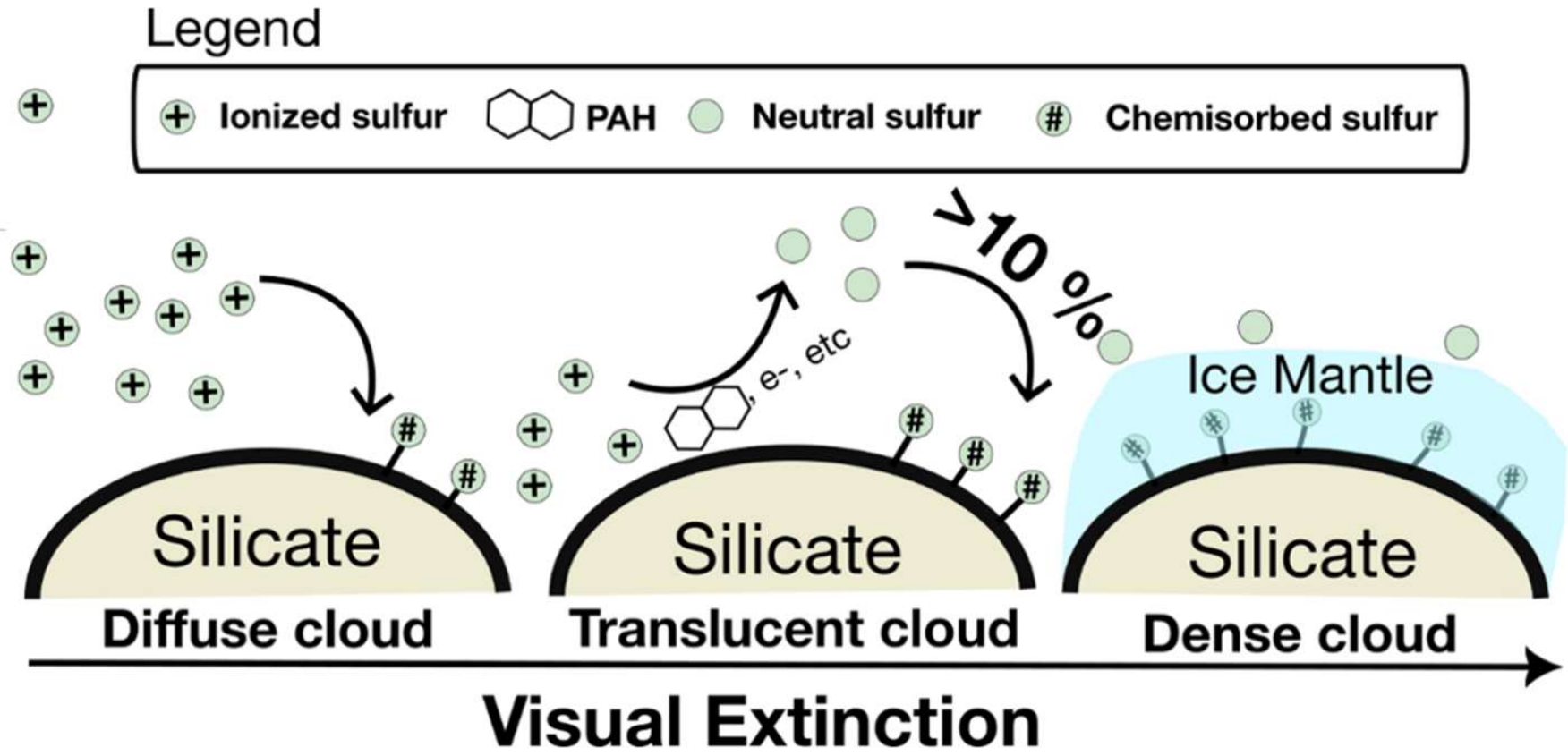

**Figure 8.** Schematic representation of the evolution of sulfur in light of our experiments. The diagram provides a conceptual overview of sulfur depletion driven by visual extinction.

This specialized entrapment framework accounts for different depletion factors for sulfur in different astronomical regions (e.g., Fuente et al. 2023). Considering that the chemisorption behavior is highly substrate-dependent and differs between silicates and carbonaceous dust (Draine 2003), the relative abundances of these dust will correlate with the sulfur depletion factor before ice mantles form. Furthermore, because chemisorbed sulfur survives after ice mantle sublimation, it remains securely locked until released by further heating, such as in shock regions. Observing the release of sulfur in the shock environments will provide direct observational evidence of the mechanism shown in this work. This observational approach has been explored by Anderson et al. (2013) using Spitzer data, and can be significantly improved with dedicated JWST observations.

Several choices in the chemical model are based on quantum chemical calculations, see Appendix B for details. Briefly, for the hydrogenation of the S atom, we found that it is highly endothermic considering the Langmuir–Hinshelwood mechanism, where it is, in principle, possible through the Eley–Rideal mechanism, depending on the spin-relaxation timescales of sulfur atom on the surface (Figure 5(b)). Subsequent hydrogenation of HS to $H_2S$ is discarded, based on the fast dissociation of $H_2S$ on nanosilicate clusters (Perrero et al. 2024b), preventing any accumulation of $H_2S$ on silicate grains.

## 5. Conclusion

In this study, we achieved the first direct, in-situ measurement of atomic sulfur on amorphous silicate and c-ASW surfaces using the PSD-REMPI method, elucidating S-atom behavior during thermal processing under conditions relevant to realistic MCs. Our results demonstrate that a significant fraction of elemental sulfur becomes securely locked onto silicate dust grains through chemisorption. By evaluating several possible pathways during heating, we identified that diffusive recombination leads to the first in-situ observation of $S_2$ formation, strongly suggesting the subsequent formation of larger sulfur clusters. Moreover, the mono-atomic thermal desorption of S atom is negligible while co-desorption with $H_2O$ sublimation is highly efficient.

The remarkable refractory nature of S and $S_2$ residues upon annealing to room temperature confirms stable chemisorption on the $Mg_2SiO_4$ surface, a finding strongly supported by recent theoretical models (Perrero et al. 2024b; Hansson et al. 2026). Simulating interstellar ice evolution through UV photolysis of ice-capped $H_2S$ (Exp. 6) confirmed that the atomic S and $S_2$ remain on silicates even after water sublimation. In contrast to previous experiments utilizing higher $H_2S$ concentrations, our current low-coverage experiments did not show the S-bearing products such as $SO_2$ and $H_2SO_4$. This indicates that the production of such species is inefficient in the realistic interstellar environment, although the formation of larger sulfur allotropes ($S_n$, $n \geq 2$) remains a viable pathway.

Contextualization our experimental results within an astrochemical modelling framework accounting for chemisorption confirms that chemisorbed sulfur indeed represents the major sulfur reservoir in MCs, driven by efficient depletion during the translucent cloud phase. We propose that sulfur depletion factors across different clouds must correlate with the fraction of silicate dust grains. Finally, we suggest that future high-sensitivity JWST observations of heavily shocked regions will serve as a critical probe to validate our hypotheses.

Acknowledgement

This work was supported by JSPS KAKENHI Grant Numbers JP22H00159, JP24K00686, JP25H00677, JP25KF0001, and JP25K07364. G.M. acknowledges support from MICIU/AEI/10.13039/501100011033 and ERDF/ESF+ (projects PID2024-156686NB-I00 and RYC2022-035442-I), CSIC (project 20245AT016), and ERC grant 101218790 (Isocosmos) funded by the European Union. Views and opinions expressed are those of the author(s) only and do not necessarily reflect those of the European Union or the ERC Executive Agency; neither can be held responsible for them. We acknowledge the computational resources provided by the DRAGO computer cluster managed by SGAI-CSIC and the Galician Supercomputing Center (CESGA). The supercomputer FinisTerrae III and its permanent data storage system have been funded by the Spanish Ministry of Science and Innovation, the Galician Government, and the European Regional Development Fund (ERDF).

## Appendix A: Binding energy of sulfur on forsterite

The contextualization of our experimental results in an astrochemical modelling accounting for chemisorption confirms that chemisorbed sulfur represents indeed the major sulfur reservoir in MC, following efficient depletion during the translucent cloud phase. We propose that sulfur depletion factors across clouds must correlate with the fraction of silicate dust and suggest future high-sensitivity JWST observations of heavily shocked regions as a probe to validate our hypotheses.

One of the key questions addressed in the main text is whether a chemisorbed S atom remains bound as is to the surface or is susceptible to further chemical processing. To investigate the fate of sulfur adsorbed on forsterite, we performed dedicated quantum chemical calculations. The forsterite surface was modeled using the nanosilicate cluster introduced by Perrero et al. (2024b), which we reoptimized at the ωB97M-D4REV level of theory (Mardirossian & Head-Gordon 2016; Müller et al. 2023) with the ma-def2-TZVP basis set (Zheng et al. 2011) using ORCA 6 (Neese 2025). All the reported energies in this Section include zero-point energy corrections. Following reoptimization, a ground-state S atom, S($^3$P), was placed on the cluster in the adsorption configuration reported by Perrero et al. (2024b) (Figure 5). From the resulting equilibrium structure, we derive a binding energy of approximately 9,700 K for the adsorbed triplet S atom. This value falls within the range (7,500-10,200 K) for nano-olivine clusters binding through the Mg sites (Perrero et al. 2024b). We note that these binding energies, both in the present work and in Perrero et al. (2024b) may be overestimated due to the enhanced deformation energy associated with finite nanocluster models compared to periodic slabs. Nevertheless, the calculated value clearly indicates strong adsorption of sulfur in the triplet state.

Starting from the optimized triplet adsorption geometry, we performed a spin-flip calculation to investigate the corresponding singlet state. We find that the singlet adsorption state is stabilized by approximately 4,760 K relative to the triplet, revealing an inversion of the energetic ordering observed for isolated gas-phase sulfur. This stabilization originates from a markedly different adsorption mode. In the triplet state, sulfur interacts mainly through a surface Mg atom, whereas in the singlet state it adopts a three-center coordination involving the Mg atom, a neighboring silicate O atom, and the S atom itself (Figure 5(a)).

**Appendix B: Hydrogenation of sulfur**

The key step governing the subsequent evolution of adsorbed sulfur is therefore the intersystem crossing (ISC) from the metastable triplet adsorption state to the lower-energy singlet state. The efficiency of this process determines whether sulfur remains trapped in the triplet configuration long enough to undergo surface reactions or instead relaxes to the more stable singlet adsorption complex. Considering relevant timescales for the ISC and the frequency of adsorption of H atoms, we have subsequently investigated the hydrogenation of sulfur in both triplet and singlet adsorption states, through the Langmuir–Hinshelwood (LH) mechanism, and the Eley–Rideal (ER) mechanism. The summary of our findings is gathered in Figure 5(b). For the LH mechanism, we find that a pre-adsorbed hydrogen atom preferentially forming a surface hydroxyl (OH) group does not subsequently add to the adsorbed sulfur atom. The overall reaction is strongly endothermic, with a reaction energy of approximately 17,200 K, making this pathway unfavorable under interstellar conditions. This behavior contrasts with the hydrogenation of adsorbed oxygen atoms reported by Molpeceres et al. (2019). The difference is nevertheless chemically intuitive: transferring a hydrogen atom from a surface hydroxyl group to sulfur requires breaking a strong O–H bond and forming a weaker S–H bond. As a result, the process is energetically disfavored, reflecting the substantially greater bond strength of hydroxyl groups relative to thiol groups. The ER addition reaction, on the other hand, is always exothermic, provided that there is no adsorption of the hydrogen atom on the surface prior to reaction. The ER addition to the pre-adsorbed triplet sulfur is found, via a relaxed scan of the potential energy surface, not only exothermic, but also barrierless, indicating that the reaction must proceed. The dissociation of the resulting HS radical has a barrier on the surface of 3,342 K, in contrast to what is found for $H_2S$ in Perrero et al. (2024b) (less than 1,200 K, see below). The same reaction in the singlet channel does not proceed (Figure 5(b)), where the adsorption of H on a neighboring O atom of the nanocluster was found to be favored. For that reason, we determine that the hydrogenation of adsorbed S is deeply linked with the ISC timescales, whose determination is beyond the scope of this work.

Finally, it remains to assess whether a hypothetical adsorbed HS radical could undergo a second hydrogenation step to form $H_2S$. Our calculations for the LH mechanism indicate that this reaction is endothermic and therefore thermodynamically non-viable. However, even if the reaction were exothermic, or if the ER mechanism provided an efficient alternative route, the formation of stable $H_2S$ on the forsterite surface would still be unlikely. Indeed, Perrero et al. (2024b) showed that adsorbed $H_2S$ is unstable on forsterite with respect to dissociation into HS and H. Their calculations showed that this dissociation process is strongly exothermic and proceeds through relatively small activation barriers, below 1,200 K (≈10 kJ $mol^{-1}$) even for the least favorable adsorption sites. Such barriers are sufficiently low that they can be easily crossed through quantum tunneling under interstellar conditions. Consequently, even if $H_2S$ was formed on the grain surface, it would rapidly dissociate back into HS and H, preventing the accumulation of molecular hydrogen sulfide on forsterite grains.

**Appendix C: Astrochemical modeling**

To estimate what fraction of sulfur can be trapped in the chemisorbed state before the silicate surface of dust grains are covered by ices, we have conducted astrochemical network simulations under translucent molecular cloud conditions as discussed in main text. We adopt the rate equation approach utilizing Rokko code (Furuya et al. 2015; Furuya 2024). We adopt the three-phase model, which includes chemical species in the gas phase, on the surface of the ice, and in the bulk ice mantle, assuming that the bulk mantle has uniform chemical composition (Hasegawa & Herbst 1993). As chemical processes, we consider gas-phase reactions, interactions between gas and (icy) grain surfaces, and surface reactions. Our chemical network is based on that developed by Garrod (2013), but the gas phase chemical network is largely replaced with those in KIDA 2024 network (Wakelam et al. 2024).

For this study, we have additionally considered chemisorbed state of H atoms, O atoms, OH, $H_2O$, S atoms, HS, and $H_2S$ in addition to their physisorbed state. Hereafter, species with a '#' suffix denote chemisorbed species. For H#, we largely follow the method by Thi et al. (2020). Briefly, H atoms can chemisorb either directly *via* the ER mechanism or after first physisorbing onto the surface. The barrier for chemisorption is assumed to be rectangular, with a height of 400 K and a width of 0.5 Å. Chemisorbed hydrogen atoms subsequently react with other hydrogen atoms *via* either the LH or ER mechanisms without an activation barrier, leading to the formation of $H_2$. We assume that O and S atoms chemisorb in a similar manner to H atoms. It is important to stress that once the surface is covered by a complete monolayer of adsorbates, regardless of whether the adsorbates are physisorbed or chemisorbed, no additional chemisorbed species are formed in our models. Any species adsorbed thereafter are therefore treated as physisorbed species. Following Molpeceres et al. (2019), we set the rate constants for the hydrogenation of O# and OH# through the LH mechanism to $10^{-14}$ s$^{-1}$, while those reactions through the ER mechanism are assumed to occur barrierlessly. The hydrogenation of S# *via* both the LH and ER mechanisms is neglected (see Appendix B). The binding energy of the chemisorbed species is set to 10,000 K, so that surface diffusion and thermal desorption are negligible in our models. We consider photodissociation of OH# and $H_2O$#, of which the rate constants are assumed to be the same as those for the corresponding physisorbed species. We also consider the UV photodesorption of the chemisorbed species, assuming the photodesorption yields per incident UV photon of $10^{-4}$.

The dust-to-gas mass ratio is set to 0.01, assuming dust size distribution follows the MRN distribution (Mathis et al. 1977) ranging from 0.005 μm to 0.25 μm. The dust size distribution is divided into four populations, and species adsorbed on different dust populations is treated as different species. For simplicity, we assume all dust grain surfaces are composed of silicates. As a result, the total number of surface binding sites on silicate surfaces is ~$7 \times 10^{-6}$ $n_{\mathrm{H}}$, indicating that ~50% of total sulfur can be trapped as the chemisorbed state on the silicate surface in our simulations.

We consider neutral and singly charged states for each dust population. In astrochemical simulations, it is often assumed that ion-grain collisions yield the same products as the corresponding gas-phase recombination reactions. We largely follow this assumption. As an exception however, we assume $S^+$ can land on dust grains as atomic S in the physisorbed state with the sticking probability of unity, following Ruffle et al. (1999). In this context, negatively charged dust grains are important, as $S^+$ is attracted to them *via* Coulomb interaction, resulting in an enhanced collision cross section compared to the geometric cross section (Cazaux et al. 2022; Fuente et al. 2023). The effect of Coulomb interaction is more significant for smaller grains (Draine & Sutin 1987). We calculate the rate constants for the collision between grains and ions or electrons following Draine and Sutin (1987). The Coulomb attraction is not relevant to oxygen, which is primarily in neutral O atoms, because of the higher ionization energy of O atoms than H atoms. The rates for the photoelectric charging and photodetachment of grains are calculated following Weingartner and Draine (2001), using parameters appropriate for silicates. The dust opacity is calculated with the dsharp_opac package (Birnstiel et al. 2018), adopting optical constants for astronomical silicates (Draine 2003).

Neutral and singly charged PAHs and the collisions with PAHs and ions/electrons are also included in our models, following Thi et al. (2020). As a typical PAH, we use a circumcoronene ($C_{54}H_{24}$) with the radius of 4.7 Å and the PAH abundance with respect to $3 \times 10^{-7}$ $n_H$. The rate constants for the collision between PAHs and ions or electrons are calculated considering Coulomb attraction (Draine & Sutin 1987). Note that we do not allow adsorption of any species on PAHs. Therefore, the inclusion of PAHs slows down the depletion of gas phase sulfur by the freeze-out onto dust grains, because PAHs neutralize $S^+$ into S atoms.

Elemental abundances are taken from EA2 set of Wakelam and Herbst (2008), which is based on the high-metal abundances observed in the rho Oph diffuse cloud, where the elemental sulfur abundance is $1.5 \times 10^{-5}$ $n_H$. Initially, all elements are in the form of atoms or atomic ions except for hydrogen, which is in $H_2$. We assume the Draine field as the interstellar radiation field, and the cosmic-ray ionization rate of $H_2$ is set to $6 \times 10^{-17}$ $s^{-1}$, which is appropriate for low extinction gas (< a few mag) (Obolentseva et al. 2024).

We run a small grid of pseudo-time dependent models, varying in the gas density (from 10 $cm^{-3}$ to 1,000 $cm^{-3}$) and the visual extinction ($A_V$; from 0.3 mag to 1.5 mag). The dust temperature is calculated using the formula by Hocuk et al. (2017). While in reality, the dust temperature depends on the size and the gas temperature can be different from the dust temperature, we assume the different dust populations and the gas have the same temperature for simplicity. Figure C1 shows the maximum fraction of sulfur trapped in S# at different density and $A_V$. Our model predicts that ~10–20 % of overall sulfur can be trapped in S# when the density is higher than 100 $cm^{-3}$. If a higher UV photodesorption yield of $10^{-3}$ is adopted for the chemisorbed species, the fraction of sulfur trapped in S# at the gas density of 100 $cm^{-3}$ decreases compared to the fiducial case, whereas the results at the density of 1,000 $cm^{-3}$ remain largely unchanged. Then, the photodesorption yield of S# would be important parameter for controlling at which density the significant depletion of gas-phase sulfur begins during the evolution from the diffuse to denser gas.

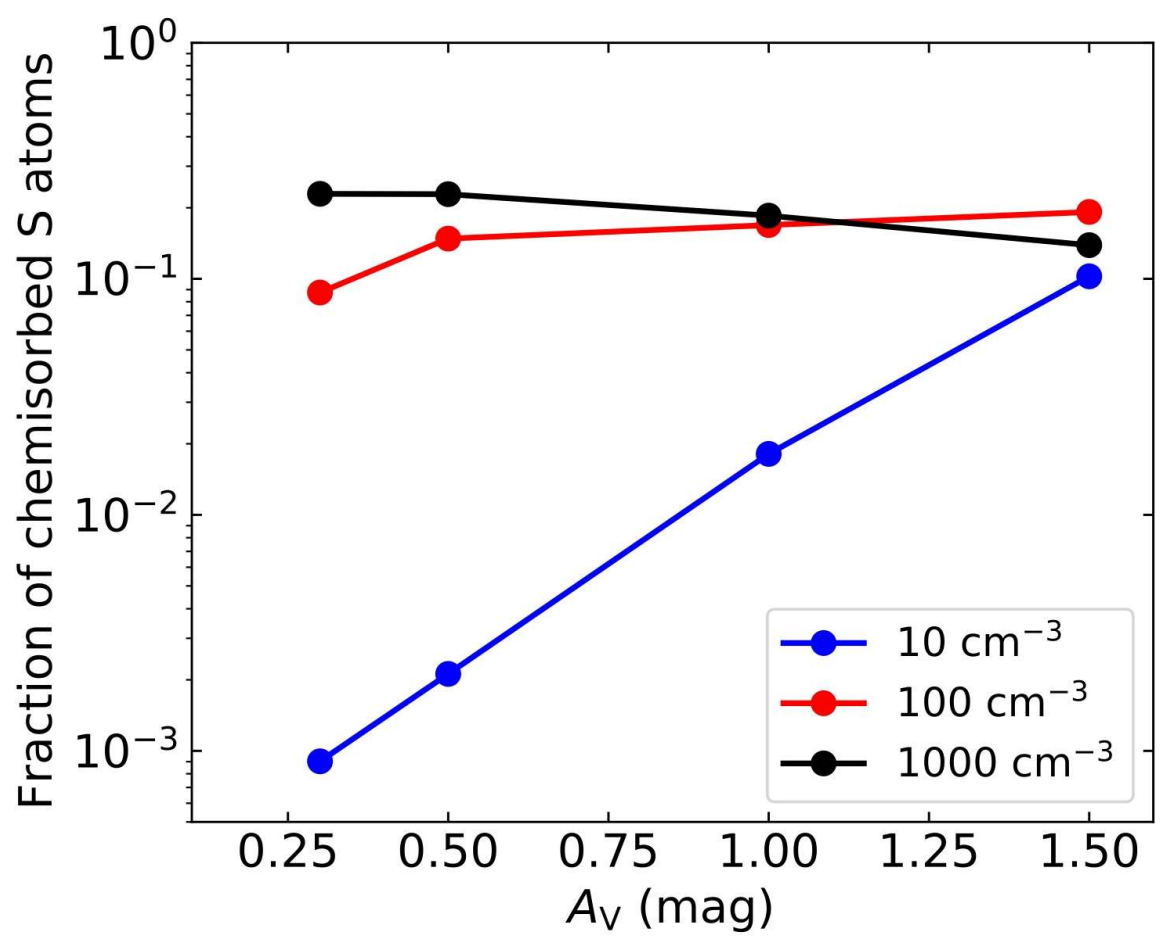

**Figure C1.** The maximum fraction of sulfur trapped in chemisorbed sulfur atoms at the different gas densities as a function of visual extinction, predicted by the astrochemical network model.

## References

Anderson, D. E., Bergin, E. A., Maret, S., & Wakelam, V. 2013, ApJ, 779, 141
Aponte, J. C., Buckner, D. K., Mojarro, A., et al. 2026, GeCoA
Bariosco, V., Pantaleone, S., Ceccarelli, C., et al. 2024, MNRAS, 531, 1371
Barnes, M., Baker, J., Dyke, J., & Richter, R. 1992, CP, 166, 229
Birnstiel, T., Dullemond, C. P., Zhu, Z., et al. 2018, ApJL, 869, L45
Boogert, A. A., Gerakines, P. A., & Whittet, D. C. 2015, ARA&A, 53, 541
Brewer, P., Van Veen, N., & Bersohn, R. 1982, CPL, 91, 126
Carrascosa, H., Muñoz Caro, G., Martín-Doménech, R., et al. 2024, MNRAS, 533, 967
Cazaux, S., Carrascosa, H., Muñoz Caro, G., et al. 2022, A&A, 657, A100
Cazaux, S., & Tielens, A. 2004, ApJ, 604, 222
Cecchi-Pestellini, C., & Aiello, S. 1992, MNRAS, 258, 125
Chen, Y.-J., Juang, K.-J., Nuevo, M., et al. 2014, ApJ, 798, 80
Coulaud, L., Wang, J., Herath, A., et al. 2025, PCCP, 27, 19324
Cuppen, H., Walsh, C., Lamberts, T., et al. 2017, SSRv, 212, 1
Das, A., Sil, M., Gorai, P., Chakrabarti, S. K., & Loison, J.-C. 2018, ApJS, 237, 9
Di Genova, G., Perrero, J., Rosi, M., et al. 2025, ACS Earth & Space Chemistry, 9, 844
Draine, B., & Sutin, B. 1987, ApJ, 320, 803
Draine, B. T. 2003, ARA&A, 41, 241
Druard, C., & Wakelam, V. 2012, MNRAS, 426, 354
Francisco, J. S., Lyons, J. R., & Williams, I. H. 2005, JCP, 123
Fuente, A., Esplugues, G., Rivière-Marichalar, P., et al. 2025, ApJL, 986, L17
Fuente, A., Rivière-Marichalar, P., Beitia-Antero, L., et al. 2023, A&A, 670, A114
Furuya, K. 2024, ApJ, 974, 115
Furuya, K., Aikawa, Y., Hincelin, U., et al. 2015, A&A, 584, A124
Furuya, K., Oba, Y., & Shimonishi, T. 2022, ApJ, 926, 171
Garrod, R. T. 2013, ApJ, 765, 60
Garrod, R. T., Wakelam, V., & Herbst, E. 2007, A&A, 467, 1103
Giustini, A., Di Genova, G., Skouteris, D., et al. 2024, ESC, 8, 2318
Hansson, K., Sameera, W., Esmerian, C. J., et al. 2026, A&A, 707, A54
Hasegawa, T. I., & Herbst, E. 1993, MNRAS, 263, 589
Heays, A., Bosman, A., & van Dishoeck, E. 2017, A&A, 602, A105
Herath, A., McAnally, M., Turner, A. M., et al. 2025, Nat Commun, 16, 5571
Hocuk, S., Szűcs, L., Caselli, P., et al. 2017, A&A, 604, A58
Jefferts, K., Penzias, A., Wilson, R., & Solomon, P. 1971, ApJ, 168, L111
Jiménez-Escobar, A., & Muñoz Caro, G. 2011, A&A, 536, A91
Kama, M., Shorttle, O., Jermyn, A. S., et al. 2019, ApJ, 885, 114
Kouchi, A., Tsuge, M., Hama, T., et al. 2021, ApJ, 918, 45
Laas, J. C., & Caselli, P. 2019, A&A, 624, A108
Lauretta, D. S., Connolly Jr, H. C., Aebersold, J. E., et al. 2024, M&PS, 59, 2453
Mardirossian, N., & Head-Gordon, M. 2016, JCP, 144
Mathis, J. S., Rumpl, W., & Nordsieck, K. H. 1977, ApJ, 217, 425
McClure, M. K., Rocha, W., Pontoppidan, K., et al. 2023, NatAs, 7, 431
Molpeceres, G., Rimola, A., Ceccarelli, C., et al. 2019, MNRAS, 482, 5389
Müller, M., Hansen, A., & Grimme, S. 2023, JCP, 158
Neese, F. 2025, WIREs Comput Mol Sci, 15, e70019

Oba, Y., Tomaru, T., Lamberts, T., Kouchi, A., & Watanabe, N. 2018, NatAs, 2, 228
Obolentseva, M., Ivlev, A., Silsbee, K., et al. 2024, ApJ, 973, 142
Okabe, H. 1978, Photochemistry of small molecules, Vol. 431 (Wiley New York)
Penzias, A., Solomon, P., Wilson, R., & Jefferts, K. 1971, ApJ, 168, L53
Perrero, J., Beitia-Antero, L., Fuente, A., Ugliengo, P., & Rimola, A. 2024a, ApJ, 971, 36
Perrero, J., Beitia-Antero, L., Fuente, A., Ugliengo, P., & Rimola, A. 2024b, MNRAS, 527, 10697
Perrero, J., Enrique-Romero, J., Ferrero, S., et al. 2022, ApJ, 938, 158
Ruffle, D., Hartquist, T., Caselli, P., & Williams, D. 1999, MNRAS, 306, 691
Shingledecker, C. N., Lamberts, T., Laas, J. C., et al. 2020, ApJ, 888, 52
Sie, N.-E., Tsuge, M., Nakai, Y., & Watanabe, N. 2024, CPL, 848, 141384
Slavicinska, K., Boogert, A., van Dishoeck, E., et al. 2025, A&A, 693, A146
Snow, T. P., & McCall, B. J. 2006, ARA&A, 44, 367
Snow, T. P., & Witt, A. N. 1996, ApJ, 468, L65
Steadman, J., & Baer, T. 1988, JCP, 89, 5507
Taillard, A., Martín-Doménech, R., Carrascosa, H., et al. 2025, A&A, 694, A263
Thi, W., Hocuk, S., Kamp, I., et al. 2020, A&A, 634, A42
Tieftrunk, A., Pineau des Forets, G., Schilke, P., & Walmsley, C. 1994, A&A, 289, 579
Tsuge, M., Hidaka, H., Kouchi, A., & Watanabe, N. 2020, ApJ, 900, 187
Tsuge, M., Molpeceres, G., Aikawa, Y., & Watanabe, N. 2023, NatAs, 7, 1351
Tsuge, M., & Watanabe, N. 2023, PJAB, 99, 103
van Dishoeck, E. F., & Black, J. H. 1989, ApJ, 340, 273
Vidal, T. H., Loison, J.-C., Jaziri, A. Y., et al. 2017, MNRAS, 469, 435
Wakelam, V., Gratier, P., Loison, J.-C., et al. 2024, A&A, 689, A63
Wakelam, V., & Herbst, E. 2008, ApJ, 680, 371
Watanabe, N., & Kouchi, A. 2008, Prog Surf Sci, 83, 439
Watanabe, N., & Tsuge, M. 2020, JPSJ, 89, 051015
Weingartner, J. C., & Draine, B. 2001, ApJS, 134, 263
Yang, X., Hua, L., & Li, A. 2024, ApJ, 974, 30
Yokoyama, T., Nagashima, K., Nakai, I., et al. 2022, Sci, 379, eabn7850
Zheng, J., Xu, X., & Truhlar, D. G. 2011, Theor Chem Acc, 128, 295